\documentclass{article}	
\usepackage{authblk}   
\usepackage{hyperref} 
\usepackage{geometry}                		
\usepackage{graphicx}				
\usepackage{amssymb}

\title{The constraints on the stochastic gravitational wave background from cosmic strings by an electromagnetic resonance system}
\author[1]{Jin Li}
\author[1]{Meijin Li}
\author[2]{Nan Yang}
\author[3]{Li Wang\textsuperscript{a}}
\author[1]{Hao Yu}
\author[1]{Yingzhou Huang}
\author[4]{Kai Lin}
\author[1]{Zi-Chao Lin}
\author[1]{Fangyu Li}

\affil[1]{College of Physics, Chongqing University, Chongqing 401331, China}
\affil[2]{Department of Electronical Information Science and Technology, Xingtai Key Laboratory for Research and Application of Robot Intelligent Detection and Sorting Technology, Xingtai University, Xingtai 054001, China}
\affil[3]{Key Laboratory of Optoelectronic Technology and Systems (Ministry of Education), Chongqing University, Chongqing 401331, China\\

$^{a}$Email: \href{mailto:wangliyu@cqu.edu.cn}{wangliyu@cqu.edu.cn}}
\affil[4]{Universidade Federal de Campina Grande, Campina Grande, PB 58429-900, Brasil\\
\textit{and} Instituto de Física, Universidade de São Paulo, São Paulo, Brasil}
\date{}							
\begin{document}

\maketitle

\begin{abstract}
As one of the primary detection targets for contemporary gravitational wave (GW) observatories, the stochastic gravitational wave background (SGWB) holds significant potential for enhancing our understanding of the early universe's formation and evolution. Studies indicate that the SGWB spectrum from cosmic strings can span an extraordinarily broad frequency range, extending from extremely low frequencies up to the microwave band. This work specifically investigates the detectability of cosmic string SGWB signals in an electromagnetic (EM) resonance system at GHz frequency. We present a systematic analysis encompassing: (1) the response of high frequency gravitational waves (HFGWs) in such EM resonance system. (2) the development and application of fundamental data processing protocols in the EM resonance system. Our results demonstrate that the EM system shows promising sensitivity to detect cosmic string SGWB signals with tension parameters $G\mu\geq 10^{-11}$ (the corresponding dimensionless amplitude $h \geq 10^{-33}$ at 1 GHz), while potentially establishing new constraints for $G\mu\leq 10^{-11}$ in the microwave band. These findings would complement existing multi-band SGWB observations and provide additional constraints on cosmic-string tension parameters in GHz frequency regimes.\\
{\bf{keywords:} }stochastic gravitational wave background (SGWB), cosmic string, electromagnetic (EM) resonance system, microwave band
\end{abstract}
\section{Introduction}
\label{intro}
The detection of the first gravitational wave (GW) event from the coalescence of a compact binary black hole system by the LIGO-Virgo Collaboration in 2015 \cite{ligo2017gravitational} marked a major milestone in the field of GW astronomy. Since then, GW astronomy has heralded an era of regular and routine observation of the compact binary coalescence (CBC) \cite{nitz20234}. In addition to those individually distinguishable CBCs detected by current ground-based detectors, a large number of CBC events with low signal-to-noise ratio (SNR) overlap and collectively form the most common type of stochastic gravitational wave background (SGWB) from astrophysical sources \cite{allen1999detecting,sathyaprakash2009physics,christensen2018stochastic,renzini2022stochastic,meacher2015mock,farmer2003gravitational}, which cannot generate gravitational waves in the GHz frequency range. The SGWB also receives contributions from cosmological sources \cite{caprini2018cosmological}, such as phase transitions \cite{witten1984cosmic,kosowsky1992gravitational,dev2016probing,von2020peccei,zhou2022gravitational}, cosmic strings \cite{siemens2007gravitational,damour2000gravitational,sarangi2002cosmic}, and inflation \cite{turner1997detectability,guzzetti2016gravitational} and other early universe phenomena. These processes can generate gravitational waves potentially observable in the GHz frequency range.

In this paper, we focus on the SGWB generated by cosmic strings. As a kind of SGWB sources, the dominant GW emission from cosmic string networks originates from loop decay during network evolution, occurring through string self intersections and collisions \cite{sousa2020full}. The cumulative effect of the emissions throughout cosmic history generates a SGWB through incoherent superposition \cite{siemens2007gravitational,damour2000gravitational,vilenkin1981gravitational,vachaspati1985gravitational,damour2005gravitational,PhysRevD.81.104028}. So the study of the SGWB is crucial for providing unique insights into the early universe, fundamental physics, and astrophysical populations and processes \cite{wang2023probing}. Detecting and characterizing the SGWB can be meaningful to validate inflation models and explore the physics of the early universe \cite{christensen2018stochastic,caprini2018cosmological,kuroyanagi2018probing}.

The SGWB from cosmic strings has an extremely wide frequency band ($10^{-18}$–$10^{10}$Hz)\cite{giovannini2009thermal,PhysRevD.67.104008}, so that it is included as a target signal in the science goals of all gravitational wave detectors, e.g., the extremely low-frequency ($\sim10^{-17}$Hz) CMB observatories (BICEP/Ali\\CPT), the nanohertz ($\sim10^{-9}$Hz) pulsar timing arrays (NANOGrav/EPTA/PPTA/CPTA), the millihertz ($\sim10^{-3}$Hz) space based detectors (LISA/Taiji/Tianqin), the ground-based detectors (LIGO/Virgo) at $10$–$10^{3}$Hz, For each detecor, the constraint on the cosmic string parameter-string tension ($G\mu$) has been estimated \cite{buchmuller2020nanograv,abbott2019search,ade2016planck,blanco2014number}. To further joint test cosmic string models across a broader frequency range, we propose to analyze the detectability of the SGWB from cosmic strings in the GHz frequency band and derive constraints on $G\mu$ in the microwave frequency band. That will complement existing limits and provide a more comprehensive probe of cosmic string scenarios.

The primary challenge in observing the stochastic gravitational-wave background (SGWB) predicted by cosmic string models within the GHz frequency band stems from its inherent weakness, characterized by extremely small strain amplitudes ($h\sim10^{-30}-10^{-34}$) and suppressed spectral energy densities ($\Omega_{{\rm{GW}}}\sim10^{-8}-10^{-14}$) as illustrated in Fig.\ref{OmegaGW} and \ref{fig:2}. These features render the SGWB in GHz-band orders of magnitude below the sensitivity thresholds of current HFGW detectors, such as microwave cavity resonators \cite{reece1982detector,pegoraro1978electromagnetic,bernard2001detector,ballantini2004electromagnetic} or pure Inverse Gertsenshtein effect \cite{fixsen2011arcade,bowman2018absorption,WeiLF2022Experimental,dai2023gravitonphoton}. Then in order to detect HFGWs with the current experimental technology, we proposed the EM resonance system combining inverse Gertsenshtein effect with Gaussian Beam (GB) electromagnetic coupling effect \cite{PhysRevD.67.104008,li2000electrodynamical,fang2002special,de2002energy,fang2003coupling,fang2004resonant,fang2005utilization,fang2007electromagnetic,li2008perturbative,li2013high,li2020electromagnetic,li2023electromagnetic}, which has been recognized as the most sensitive and promising GW detection scheme in microwave frequency band \cite{franciolini2022hunt,ringwald2021gravitational,ashoorioon2014gravitational,giovannini2023invisible,giovannini2019blue,gehrman2023primordial,bernal2024probing}. 
A detailed description of the EM resonance system will be provided in Sect.\ref{subsec:3.1}. Building upon previous studies of this system, our work further incorporates Gaussian beams (GB) with bandwidth and accounts for dominant noise sources (e.g., background, thermal and shot noise), thereby proposing a comprehensive data processing pipeline for the system. Those ensure that the results are more closely aligned with experimental realities, bridging the gap between theoretical models and practical feasibility.

The outline of this paper is the following: In Sect.\ref{sec:2} we describe the calculation of the SGWB spectrum varying with cosmic strings tension, as well as the values of relevant cosmological parameters. In Sect.\ref{sec:3} we provide a detailed explanation of the fundamental principles and architecture of the EM resonance system, as well as an analysis of the effects induced by the SGWB from cosmic strings in the EM system. Subsequently, in Sect. \ref{sec:4} we establish a high-frequency gravitational wave (HFGW) data processing pipeline tailored to the EM resonance system, and quantify the system’s sensitivity and evaluate the detectability of the SGWB under this framework, deriving constraints on the source parameter (e.g., string tension $G\mu$). Finally, some summaries and future research directions are presented in Sect.\ref{sec:5}.  
\section{The spectrum of SGWB from cosmic strings}
\label{sec:2}  
The characterization of the stochastic gravitational wave background (SGWB) from cosmic strings typically requires numerical simulations. However, the numerical approach can become computationally expensive and time consuming due to the comprehensive exploration of the parameter space. An full analytical approximation for the SGWB spectrum produced by cosmic string networks in the standard cosmological model was derived in \cite{sousa2020full}, providing a computationally efficient alternative to full numerical simulations across the entire frequency range. So in this paper, we will use these analytical approximation formulas for the SGWB simulation, which include the fundamental mode and higher harmonic modes SGWB spectrum. Furthermore, we also considered the impact of the change of the number of relativistic species as the universe expands on the spectrum \cite{Sanidas2013,Binetruy2012}.

The SGWB we study is constrained with one free parameter: the cosmic string tensor $G\mu$, which characterizes the size of the loop with a constant $\alpha$ that we set $\alpha=0.1$. We set the total power of cosmic string emission as $\Gamma=50$. Based on Planck 2018 data \cite{aghanim2020planck}, we set $H_{0}=67.8$km/s/Mpc, and assume a flat universe where radiation and matter density parameters at the present time are $\Omega_{m}=0.308$ and $\Omega_{r}=9.1476\times10^{-5}$, respectively. Considering that cosmic string loops arises from three distinct periods: loops formed and decayed during the radiation period, loops formed during the radiation period and decayed during the matter period, and loops formed during the matter period. The corresponding SGWB energy spectrums can be calculated separately.

{\bf{Fundamental mode:}}

(i) For loops formed and decayed in the radiation region, the form of stochastic gravitational wave background is given by 
\begin{eqnarray}
\Omega_{{\rm{GW}}}^{{r}}(f) &=& \frac{128}{9}\pi A_{{r}}\Omega_{{r}}\frac{G\mu}{\varepsilon_{{r}}} \nonumber \\
&& \times \left[ \left( \frac{f(1+\varepsilon_r)}{ \frac{B_{{r}}\Omega_m}{\Omega_{{r}}} + f } \right)^{\frac{3}{2}} - \left( \frac{f(\varepsilon_r + 1)}{B_i + f} \right)^{\frac{3}{2}} \right], \label{GWr}
\end{eqnarray}
where
\begin{eqnarray*}
    \varepsilon_{{r}} = \frac{\alpha}{\Gamma G\mu}, & \quad &
    A_{{r}} = \frac{\tilde{c}}{\sqrt{2}}F\frac{v_{{r}}}{\xi_{{r}}^{3}}, \\ 
    B_{{r}} = \frac{2H_{0}\Omega_{r}^{\frac{1}{2}}}{\nu_{r}\Gamma G\mu}, & \quad &
    B_{i} = \frac{2}{\Gamma} \sqrt{\frac{2H_{0}\Omega_{r}^{\frac{1}{2}}}{t_{\rm pl}(\varepsilon_{r}+1)}},
\end{eqnarray*}
 The label $r$ indicates the radiation era and in these equations ${{v}_{r}}=0.662$, ${{\xi }_{r}}=0.271$, ${{\nu }_{r}}=1/2$, $F=0.1$, $t_{\rm pl}=10^{-43}$ s and $\tilde{c}$ is a phenomenological parameter which can be set as $\tilde{c}=0.23\pm 0.04$ \cite{martins2002extending}. 

 (ii) For loops formed in the radiation region and decayed in the matter region, their contribution to SGWB has the following form
\begin{eqnarray}
\Omega_{{\rm{GW}}}^{{rm}}(f)=32\sqrt{3}\pi \left({\Omega_{m}\Omega_{r}}\right)^{\frac{3}{4}} H_{0} \frac{A_{r}}{\Gamma} 
\frac{\left(1+\varepsilon_{r}\right)^{\frac{3}{2}}}{f^{\frac{1}{2}}\varepsilon_{r}} \nonumber \\
\quad \times \left\{ 
\frac{\left(\frac{\Omega_{m}}{\Omega_{r}}\right)^{\frac{1}{4}}}
{\left(B_{m}\left(\frac{\Omega_{m}}{\Omega_{r}}\right)^{\frac{1}{2}} + f\right)^{\frac{1}{2}}}
\left[ 2 + \frac{f}{B_{m}\left(\frac{\Omega_{m}}{\Omega_{r}}\right)^{\frac{1}{2}} + f} \right] \right. \nonumber \\
\quad \left. - \frac{1}{\left(B_{m} + f\right)^{\frac{1}{2}}} 
\left[ 2 + \frac{f}{B_{m} + f} \right] \right\},
\end{eqnarray}
where
\begin{eqnarray*}
{{B}_{m}}=\frac{3{{H}_{0}}\Omega_{m}^{\frac{1}{2}}}{\Gamma G\mu}.
\end{eqnarray*}

(iii) The contribution of loops generated in the matter period to the SGWB generation by cosmic strings is given by
\begin{eqnarray}
\Omega_{{\rm{GW}}}^{m}(f) &=& 54\pi H_{0} \Omega_{m}^{\frac{3}{2}} \frac{A_{m}}{\Gamma} \frac{\varepsilon_{m} + 1}{\varepsilon_{m}} \frac{B_{m}}{f} \nonumber \\
&& \times \left\{ \frac{2B_{m} + f}{B_{m}(B_{m} + f)} - \frac{1}{f} \frac{2\varepsilon_{m} + 1}{\varepsilon_{m}(\varepsilon_{m} + 1)} \right. \nonumber \\
&& \quad \left. + \frac{2}{f} \log \left( \frac{\varepsilon_{m} + 1}{\varepsilon_{m}} \frac{B_{m}}{B_{m} + f} \right) \right\},
\end{eqnarray}
where
\begin{eqnarray*}    
{{\varepsilon }_{m}}=\frac{\alpha }{\Gamma G\mu},{{A}_{m}}=\frac{{\tilde{c}}}{\sqrt{2}}F\frac{{{v}_{m}}}{{{\xi }_{m}}^{3}}.
\end{eqnarray*}
The label $m$ indicates the matter era and in these equations ${{v}_{m}}=0.583$, ${{\xi}_{m}}=0.625$. 

Therefore, the SGWB generated by cosmic strings can be well approximated as
\begin{eqnarray}\label{Omegaj1}
{{\Omega}^{1}_{{\rm{GW}}}}\left( f \right)=\Omega _{{\rm{GW}}}^{r}\left( f \right)+\Omega _{{\rm{GW}}}^{rm}\left( f \right)+\Omega _{{\rm{GW}}}^{m}\left( f \right),
\end{eqnarray}
for $\alpha\geq\Gamma G\mu$ \cite{sousa2020full,flauger2021improved,burden1985gravitational}.

{\bf{Higher harmonic modes:}} 

Our above analysis has focused on the fundamental emission mode ($j = 1$, i.e.,Eq.(\ref{Omegaj1})), but cosmic string loops in fact radiate gravitational waves across all harmonic modes. The total gravitational wave spectrum can be expressed as:
\begin{equation}
{{\Omega}^{*}_{{\rm{GW}}}}\left( f \right) = \sum_{j=1}^{n_*} \frac{j^{-q}}{\mathcal{E}} \Omega _{{\rm{GW}}}^{1}\left( f/j \right),
\end{equation}
where $q=4/3$, $5/3$, or $2$ for loops with cusps, kinks, or kink-kink collisions, respectively, and $\mathcal{E} = \sum_{m=1}^{n_*} m^{-q}$ is the normalization factor \cite{sousa2020full}. The inclusion of higher harmonics ($n_*\gg1$) extends the detectable SGWB to higher frequencies while slightly modifying the spectral shape.

Critically, the amplitude of the spectrum is further suppressed at high frequencies due to changes in the number of relativistic species $g_*$ during cosmic evolution. As the universe cools, particle annihilations transfer entropy to remaining relativistic species, reducing ${{\Omega }_{{\rm{GW}}}}\left( f \right)$ by a factor $(g_*/g_*')^{1/3}$ \cite{Sanidas2013}. In GHz frequencies (probing early times when $g_*' = 106.75$) this suppression becomes significant, yielding
\begin{equation}
\Omega_{{\rm{GW}}}\left( f \right) = {{\Omega }^{*}_{{\rm{GW}}}}\left( f \right) \left(\frac{g_*}{g_*'}\right)^{1/3},
\end{equation}
where $g_*$ drops to 3.36 after electron-positron annihilation and neutrino decoupling \cite{Binetruy2012}. This correction reduces the expected signal amplitude by the $\sim$1 order of magnitude in the GHz range, directly impacting the detector's sensitivity to $G\mu$.

\begin{figure}
\centering
  \includegraphics[width=0.5\textwidth]{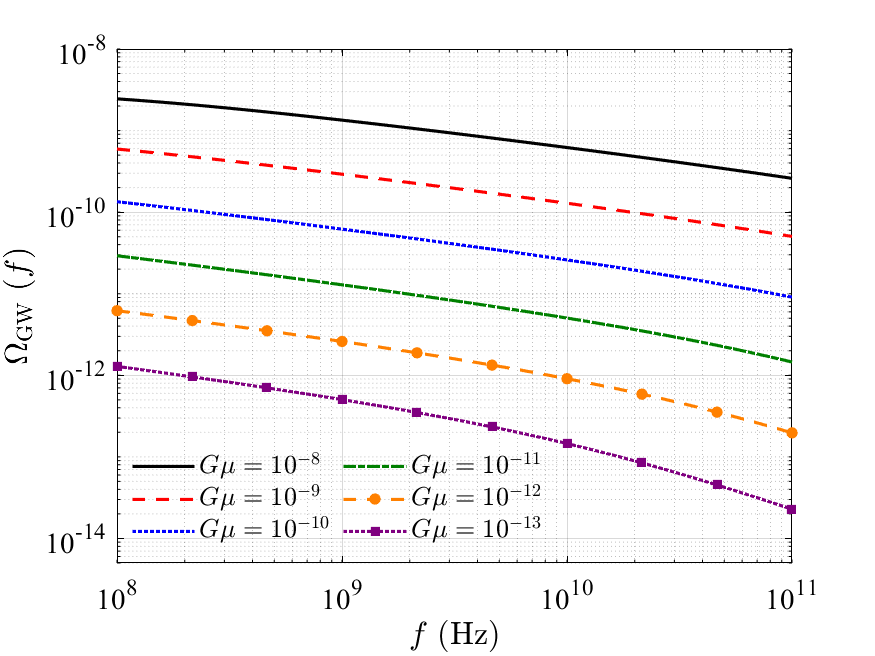}
\caption{The spectrum of the SGWB from cosmic string for $\alpha=0.1$, $\Gamma=50$, $n_*=10^{5}$, $q=4/3$, $g_*' = 106.75$, and $g_* = 3.36$, with different values of $G\mu$. The calculation is based on \cite{sousa2020full,wang2023probing} and the theoretical framework presented in Section~\ref{sec:2}.}
\label{OmegaGW}       
\end{figure}

Based on above equations, Fig.\ref{OmegaGW} shows the spectrum of the SGWB $\Omega_{{\rm{GW}}}$ in the high-frequency range with $n_*=10^{5}$, $q=4/3$.

\section{The response of SGWB from cosmic strings in HFGW electromagnetic resonance system}\label{sec:3}
As indicated in Fig.\ref{OmegaGW}, the SGWB from cosmic string exhibits potential spectral extension into the GHz frequency band – a regime fully accessible to the HFGW electromagnetic resonance system. That establishes the technical feasibility of employing such system to detect or constrain the characteristic parameters of SGWB sources. Notably, the response of HFGW in the electromagnetic resonance system can be considered as a kind of GW electromagnetic counterparts in the GHz frequency band \cite{li2023electromagnetic}. In Section \ref{subsec:3.1}, we will briefly explain the foundational physical principles governing this transduction process. Building upon this framework, Section \ref{subsec:3.2} will present a quantitative analysis of the predicted SGWB in the HFGW electromagnetic resonance architectures.

\subsection{Principle of the HFGW electromagnetic resonance system}
\label{subsec:3.1}
\begin{figure}
\centering
  \includegraphics[width=0.4\textwidth]{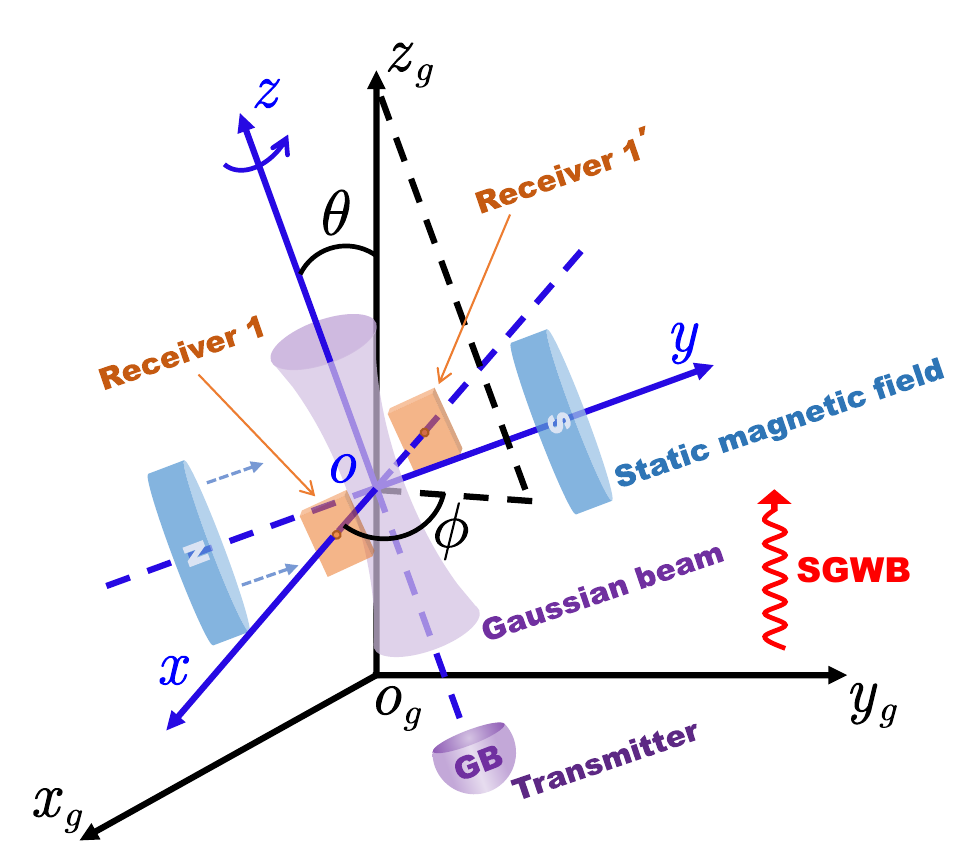}
\caption{Structure of the EM Resonance System. The system is defined in a laboratory frame with x-y-z coordinate, which describes the physical configuration of Gaussian beam (GB), static magnetic field and microwave photon receivers. Specifically, GB from the transmitter propagates along z-axis towards +z direction. Static magnetic field pointing along the y-axis is localized in the region $-l_{1}\leq z\leq l_{1}$. Microwave photon receiver 1 positioned on the +x axis, detecting photons at its location, while microwave photon receiver 1' positioned on the -x axis, symmetrically opposite to receiver 1. The effective detection area of each receiver is given by $\Delta s=\Delta z\Delta y$.}
\label{fig:1}       
\end{figure}

Generally, we consider the background geometry taken to be conformally flat being consistent with the $\Lambda$CDM paradigm, i.e.\begin{equation}\label{metric}
ds^2=\bar{g}_{\mu\nu}dx^{\mu}dx^{\nu}, ~~\bar{g}_{\mu\nu}=a^{2}(\tau)\eta_{\mu\nu},
\end{equation}
where $\eta_{\mu\nu}$ is the Minkowski metric with signature mostly positive, i.e. (-,+,+,+), $a(\tau)$ is the cosmology scale factor with a normalized value at the present time, i.e., $a(t_{0})$=1, $\tau$ is the conformal time. When GW exists in the spacetime background, it will generate perturbations on the background metric. In general relativity, these perturbations have two independent polarization modes ($\oplus$ and $\otimes$). Each polarization component $h_{ij}$ can be expressed as \cite{giovannini2009thermal,giovannini1999production,giovannini1999spikes,grishchuk1991spectra}: 
\begin{equation}\label{h}
h_{ij}=\frac{\mathcal{F}(\tau)}{a(\tau)}e^{{\rm{i}}\vec{k}\cdot\vec{x}}e_{ij},
\end{equation}
where $\vec{k}$ is the wave vector of GW, $\mathcal{F}(\tau)$ is the (complex) mode function in the universe evolution obeying \cite{li2008perturbative}
\begin{equation}\label{F}
\frac{d^{2}{\mathcal{F}}}{d\tau^2}+(k^{2}-\frac{1}{a}\frac{d^{2}{a}}{d\tau^2})\mathcal{F}=0,
\end{equation}
where $k=|\vec{k}|$. In the different evolution stages of the universe, the solution of Eq.(\ref{F}) is different. In GHz frequency domain $k^{2}\gg\frac{1}{a}\frac{d^{2}{a}}{d\tau^2}$, after neglecting $\frac{1}{a}\frac{d^{2}{a}}{d\tau^2}$ Eq.(\ref{F}) can have the general periodic solution as 
\begin{equation}\label{FH}
\mathcal{F}(\tau)=A(k)e^{-{\rm{i}}k\tau}+B(k)e^{{\rm{i}}k\tau},
\end{equation}
where $A(k)$ and $B(k)$ are the constants of integration. For the electromagnetic resonance system in the laboratory frame, we should use the intervals of the laboratory time ($t$) satisfying $cdt=a(\tau)d\tau$. So Eq.(\ref{h}) can be derived as
\begin{equation}
h_{ij}=[\frac{A(k)}{a(t)}e^{{\rm{i}}(\vec{k}\cdot\vec{x}-\omega t)}+\frac{B(k)}{a(t)}e^{{\rm{i}}(\vec{k}\cdot\vec{x}+\omega t)}]e_{ij},
\end{equation}
which can be considered as the approximate form of each monochromatic component of the SGWB in GHz band. In our study, we will be focused on a circular polarized monochromatic component of the SGWB in its propagation direction ($+z_{g}$ axis), i.e.,
\begin{equation}\label{hij1}
h_{\oplus}=A_{\oplus}e^{{\rm{i}}(k z_{g}-\omega t)}, \\
\end{equation} 
\begin{equation}\label{hij2}
h_{\otimes}={\rm{i}}A_{\otimes}e^{{\rm{i}}(k z_{g}-\omega t)}, 
\end{equation} 
 where $A_{\oplus}, A_{\otimes}$ set to be $A(k)/a(t)$, which are the dimensionless amplitudes of the SGWB in the standard GW frame system ($x_{g},y_{g},z_{g}$), and $k, \omega$ are the corresponding wave vector and angular frequency in the standard GW frame system. Suppose that the laboratory frame system of the electromagnetic resonance system($x, y, z$) is rotated by angles ($\theta, \phi$) from the standard GW frame system as Fig.\ref{fig:1}. 
Given the existence of the SGWB in all directions on cosmological scales and combined with theoretical analyses from prior studies \cite{li2008perturbative,li2023electromagnetic,tong2008using}, the response of the HFGWs system reaches its theoretical maximum when the propagation direction of the SGWB  (the $z_{g}$-direction in the standard GW frame system) aligns perfectly with the reference z-axis of the laboratory frame system(i.e., $\theta=0$ in Fig.\ref{fig:1}). Fortunately, the spatial orientation of the electromagnetic resonance system can be experimentally optimized to achieve effective alignment between the propagation direction and the system's reference axis, thereby ensuring the fulfillment of this optimal observational condition. Then according to Eqs.($\ref{hij1}$), ($\ref{hij2}$) and Eq.(4) of \cite{li2023electromagnetic} the polarization states $h_{ij}$ in the standard laboratory frame system should be:
\begin{equation}\label{h11,h22}
h_{11}=-h_{22}=h_{\oplus}\cos2\phi+h_{\otimes}\sin2\phi, \nonumber
\end{equation}
\begin{equation}\label{h12,h21}
h_{12}=h_{21}=-h_{\oplus}\sin2\phi+h_{\otimes}\cos2\phi,
\end{equation}
and other components of $h_{\mu\nu}$ are zero. Then the spacetime metric in Cartesian coordinates is $g_{\mu\nu}=\eta_{\mu\nu}+h_{\mu\nu}$. And solving the electrodynamics equations in vacuum curved spacetime
\begin{equation}\label{curveeq1}
\frac{\partial}{\partial x^{\nu}}(\sqrt{-g}g^{\mu\alpha}g^{\nu\beta}F_{\alpha\beta})=0,
\end{equation}
\begin{equation}\label{curveeq2}
\nabla_{\mu}F_{\nu\alpha}+\nabla_{\nu}F_{\alpha\mu}+\nabla_{\alpha}F_{\mu\nu}=0,
\end{equation}
where $F_{\alpha\beta}=F^{(0)}_{\alpha\beta}+F^{(1)}_{\alpha\beta}$ is the total EM field tensor composed of the background EM field tensor $F^{(0)}_{\alpha\beta}$ and the first-oder perturbative EM field tensor $F^{(1)}_{\alpha\beta}$ due to the GW. We can get the components of the perturbative EM field (in SI units) generated by the EM response to the SGWB, which are \cite{li2008perturbative,li2023electromagnetic}
\begin{equation}\label{EM11}
E^{(1)}_{x}=cB^{(1)}_{y}=\frac{{\rm{i}}}{2}kc(z+l_{1})h_{22}\hat{B}^{(0)}_{y},
\end{equation}
\begin{equation}\label{EM12}
E^{(1)}_{y}=-cB^{(1)}_{x}=\frac{{\rm{i}}}{2}kc(z+l_{1})h_{12}\hat{B}^{(0)}_{y},
\end{equation}
\begin{equation}\label{EM13}
E^{(1)}_{z}=B^{(1)}_{z}=0,
\end{equation}
where $c$ is the speed of light in a vacuum, $\hat{B}^{(0)}_{y}$ is the static magnetic field in Fig.\ref{fig:1}, pointing along the y-axis. From Eqs.(\ref{EM11}) and (\ref{EM12}), we can see that the stochastic gravitational wave background and the artificially applied transverse static magnetic field $\hat{B}^{(0)}_{y}$ give rise to a first-order perturbed electromagnetic field ($\vec{E}^{(1)}$, $\vec{B}^{(1)}$). The physically observable quantity of this perturbed field is the energy flux $\vec{S}^{(2)}=\vec{E}^{(1)}\times\vec{B}^{(1)}/\mu_{0}$ ($\mu_{0}$ is vacuum magnetic permeability), and detecting this energy flux would allow us to observe the electromagnetic effects induced by the stochastic gravitational wave background. However, this energy flux $|\vec{S}^{(2)}|\sim h_{\oplus,\otimes}^{2}$, which is extremely weak, making the detection remarkably difficult.
\begin{figure*}
\centering
  \includegraphics[width=0.4\textwidth]{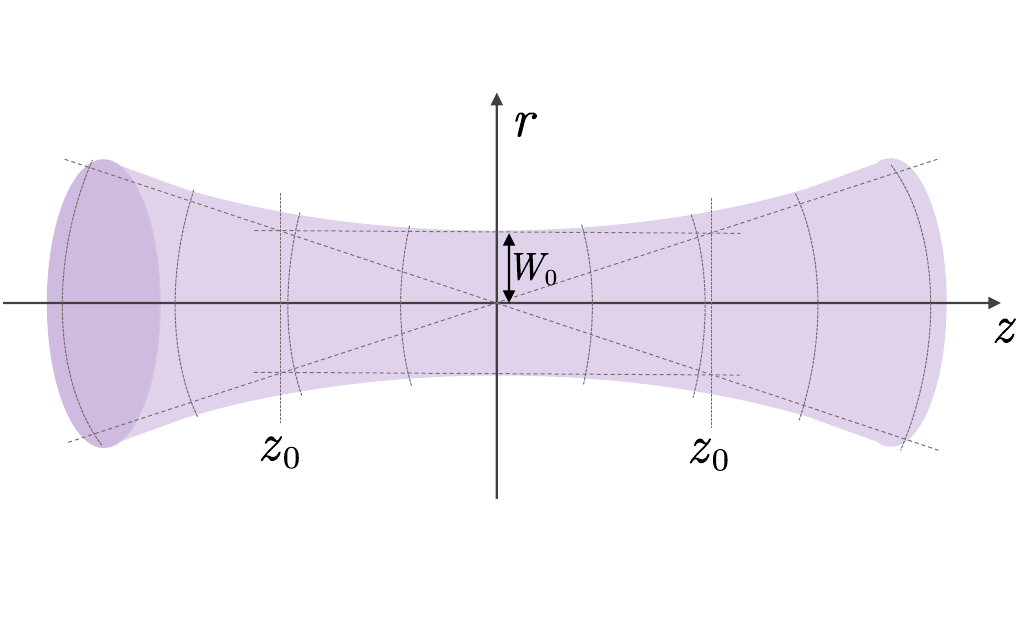}
\includegraphics[width=0.4\textwidth]{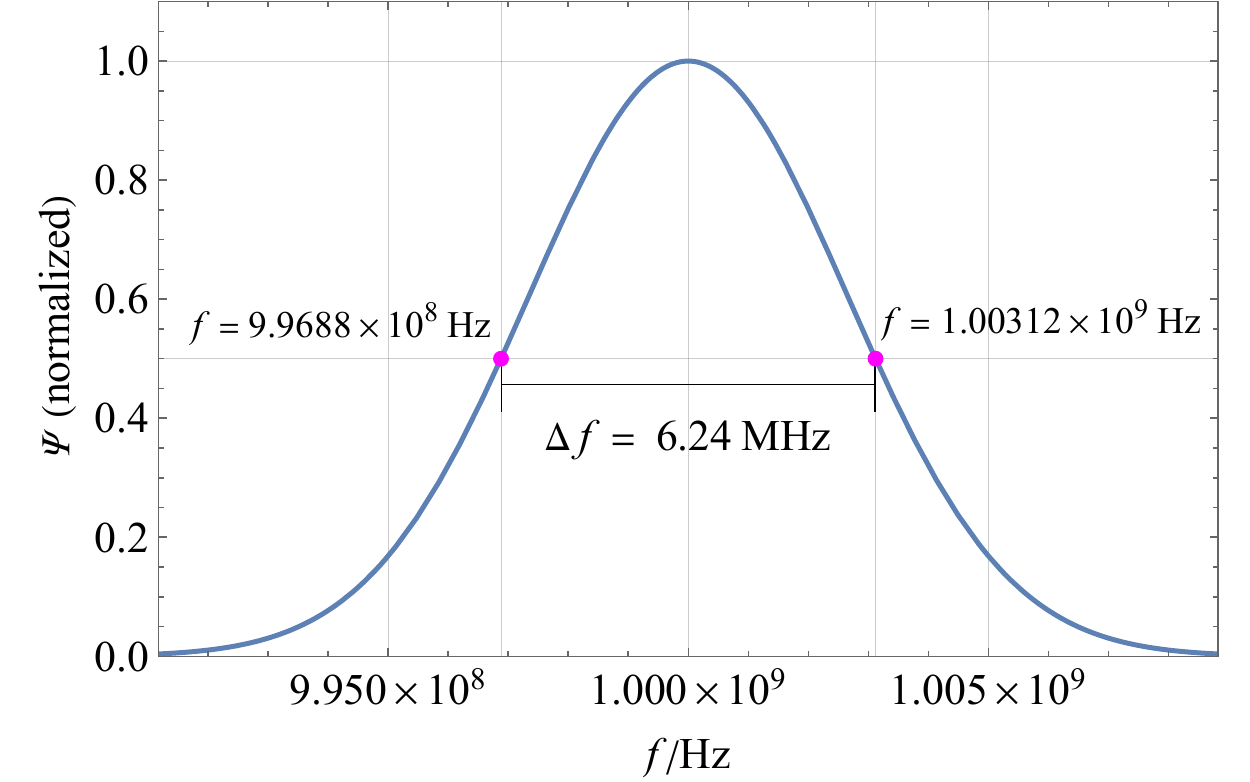}
\caption{\textbf{Left-panel:} Schematic of a Gaussian beam propagating along the z-axis, showing the beam waist $W_{0}$ (minimum radius) and the Rayleigh range $z_{0}=\pi W^{2}_{0}/\lambda_{e}$. \textbf{Right-panel:} Normalized amplitude profile of the Gaussian beam versus frequency, centered at $f_{0} = 1 \times 10^9$ Hz. The bandwidth $\Delta f = 6.24$ MHz corresponds to the full width at half maximum (FWHM).}
\label{GB}       
\end{figure*}
In order to enlarge the energy flux, we adopt a Guassian Beam (GB) $\Psi$ propagating along z-axis (see Fig.\ref{GB}), which can support a background EM field ($\vec{E}^{(0)}$, $\vec{B}^{(0)}$) and the corresponding $\vec{S}^{(0)}=\vec{E}^{(0)}\times\vec{B}^{(0)}/\mu_{0}$. Without loss of generality, we set  
\begin{equation}\label{Ex0,Ez0}
E^{(0)}_{x}=\Psi, E^{(0)}_{z}=0.
\end{equation}
where 
\begin{equation}\label{Psi}
   \Psi=\frac{\Psi_{0}e^{-(\frac{r^2}{W^2}+a_{g}^2\frac{t^2}{T^2_{p}})}}{\sqrt{1+(\frac{z}{z_{0}})^2}}e^{[{\rm{i}}(k_{e}z-\omega_{e}t-tan^{-1}\frac{z}{z_{0}}+\frac{k_{e}r^2}{2R}+\delta_{0})]},
\end{equation}
$\Psi_{0}$ is the maximum amplitude of the Gaussian beam
along the propagation direction,which can be derived from GB power ($P$) as $\Psi_{0}=\sqrt{4P\mu_{0}c/\pi W^{2}_{0}}$, $r^2=x^2+y^2$, $W=W_{0}(1+(z/z_{0})^2)^{1/2}$ ($W_{0}$ is the waist radius of the GB), the pulse width $T_p$ scales inversely with the bandwidth $\Delta f$ ($T_p \propto 1/\Delta f$), $z_{0}=\pi W^{2}_{0}/\lambda_{e}$ ($\lambda_{e}$ is the wavelength of the GB), $k_{e}=2\pi/\lambda_{e}$, $R=z+z
^{2}_{0}/z$ is the curvature radius of the wave fronts of the GB at $z$, $\omega_{e}$ is the angular frequency of the GB, $\delta_{0}$ is the phase difference between the GB and the SGWB \cite{wu2023simulation,peng2007fields}. The detailed values of GB parameters are list in Tab.\ref{tab:1}.
\begin{table}
\centering
\caption{The values of the EM system parameters. All of the parameters are chosen to exhibit values that can be realized in the proposed laboratory experiment.}
\label{tab:1}       
\begin{tabular}{ll}
\hline\noalign{\smallskip}
Parameter Name & Value  \\
\noalign{\smallskip}\hline\noalign{\smallskip}
Center frequency ($f_{0}$)  & $1 \times 10^9$ Hz \\
Beam waist radius ($W_{0}$) & 0.05 m \\
Initial phase ($\delta_0$) & $1.23\pi$ \\ 
Power ($P$) & 10 W \\ 
Bandwidth constant ($a_{g}$) & $(2\ln2)^{1/2}$ \\ 
Pulse width ($T_{p}$) & $10^{-7}$ s \\ 
Bandwidth ($\Delta f$) & 6.24 MHz \\ 
Single pulse duration ($5T_p$) & $5 \times 10^{-7}$ s \\ 
Longitudinal size of static magnetic field ($l_{1}$) & 5.7{\rm{m}} \\
Strength of static magnetic field ($\hat{B}^{(0)}_{y}$) & 3{\rm{T}} \\
\noalign{\smallskip}\hline
\end{tabular}
\end{table}
The other EM components of the background EM field can be derived from $\nabla\cdot\vec{E^{(0)}}=0$ and $\vec{B}^{(0)}=-{\rm{i}}(\nabla\times\vec{E}^{(0)})/\omega_{e}$ as
\begin{equation}\label{Ey0}
E^{(0)}_{y}=2x(\frac{1}{W^2}-\frac{ik_{e}}{2R})\int E^{(0)}_{x}dy,
\end{equation}
\begin{equation}\label{Bx0,By0}
B^{(0)}_{x}=\frac{{\rm{i}}}{\omega_{e}}\frac{\partial E^{(0)}_{y}}{\partial z}, B^{(0)}_{y}=-\frac{{\rm{i}}}{\omega_{e}}\frac{\partial E^{(0)}_{x}}{\partial z},
\end{equation}
\begin{equation}\label{Bz0}
B^{(0)}_{z}=\frac{{\rm{i}}}{\omega_{e}}(\frac{\partial E^{(0)}_{x}}{\partial y}-\frac{\partial E^{(0)}_{y}}{\partial x}),
\end{equation}

The EM field of GB ($\vec{E}^{(0)}$, $\vec{B}^{(0)}$) can interact with the first-order perturbed EM field ($\vec{E}^{(1)}$, $\vec{B}^{(1)}$) to generate an extra energy flux $\vec{S}^{(1)}=(\vec{E}^{(1)}\times\vec{B}^{(0)}+\vec{E}^{(0)}\times\vec{B}^{(1)})/\mu_{0}$ when $\omega=\omega_{e}$. In this case, our detecting object is the first-order perturbed energy flux $|\vec{S}^{(1)}|$, which represents our approach as an innovative approach distinct from current gravitational wave detection schemes in the microwave band \cite{Aggarwal2021}. Combining Eqs.(\ref{EM11})-(\ref{Bz0}), the corresponding first-order perturbed photon fluxes on the microwave receiver in the standard laboratory frame system can be calculated as
\begin{eqnarray}\label{Nx1}
N^{(1)}_{x}(x,t,\phi)=\int^{t+\Delta t}_{t}\int\int_{\Delta s}n_{x}^{(1)}dydzdt'.
\end{eqnarray}
where $n_{x}^{(1)}=<E^{(1)}_{y}B^{(0)}_{z}-E^{(0)}_{z}B^{(1)}_{y}>/\mu_{0}\hbar\omega_{e}$ is the first-order perturbed photon fluxes density in x-axis, $\Delta t$ is the signal accumulation time, $\Delta s$ is the  surface area of the receiver probe. Similarly, we can obtain
\begin{eqnarray}\label{Ny1}
N^{(1)}_{y}=\int^{t+\Delta t}_{t}\int\int_{\Delta s}\frac{1}{\mu_{0}\hbar\omega_{e}}<E^{(0)}_{z}B^{(1)}_{x}\nonumber \\
-E^{(1)}_{x}B^{(0)}_{z}>dydzdt',
\end{eqnarray}
\begin{eqnarray}\label{Nz1}
N^{(1)}_{z}=\int^{t+\Delta t}_{t}\int\int_{\Delta s}\frac{1}{\mu_{0}\hbar\omega_{e}}<E^{(1)}_{x}B^{(0)}_{y}-E^{(1)}_{y}B^{(0)}_{x}\nonumber \\
+E^{(0)}_{x}B^{(1)}_{y}-E^{(0)}_{y}B^{(1)}_{x}>dydzdt'.
\end{eqnarray}
Eqs.(\ref{Nx1})-(\ref{Nz1}) can be regarded as the interference terms between the GB and the first-order perturbed EM field induced by the HFGW. Here $<>$ means the average over a period of time. Our previous works \cite{PhysRevD.67.104008,li2008perturbative,PhysRevD.80.064013} have indicated that the perturbative first-order photon flux in the z direction ($N^{(1)}_{z}$) is completely overwhelmed by the background photon flux ($N^{(0)}_{z}\sim S^{(0)}_{z}/\hbar\omega_{e}$) in almost all of the regions, since they have the same physical distributions while $|N^{(1)}_{z}|\ll|N^{(0)}_{z}|$ (cf. Fig. 5 of \cite{PhysRevD.80.064013}). On the transverse directions ($x,y$ - axis), the perturbative first-order photon fluxes $N_{x}^{(1)}$, $N_{y}^{(1)}$ have different rates of decay compared with the transverse background photon fluxes ($N_{x}^{(0)}\sim S^{(0)}_{x}/\hbar\omega_{e}$, $N_{y}^{(0)}\sim S^{(0)}_{y}/\hbar\omega_{e}$) and their propagation directions are respectively opposite to $N_{x}^{(0)}$, $N_{y}^{(0)}$ in some
local regions. So in transverse directions, the perturbative EM field due to the HFGW is expected to be distinguishable from the background EM field. Considering $N^{(1)}_{y}$ is along the same direction as the static magnetic system's orientation, to avoid interfering with the static magnetic field distribution, it is not recommended to install photon counters on the y-axis to detect $N^{(1)}_{y}$. Therefore the transverse first-order photon fluxes (PPFs, i.e.,$N^{(1)}_{x}$) are chosen to be our signal due to the HFGW. In the following context, we will focus on the detectability of PPFs ($N^{(1)}_{x}$) from the cosmic string SGWB. 


\subsection{The signal of the SGWB from cosmic strings in the system}
\label{subsec:3.2}
\begin{figure}
\centering
  \includegraphics[width=0.5\textwidth]{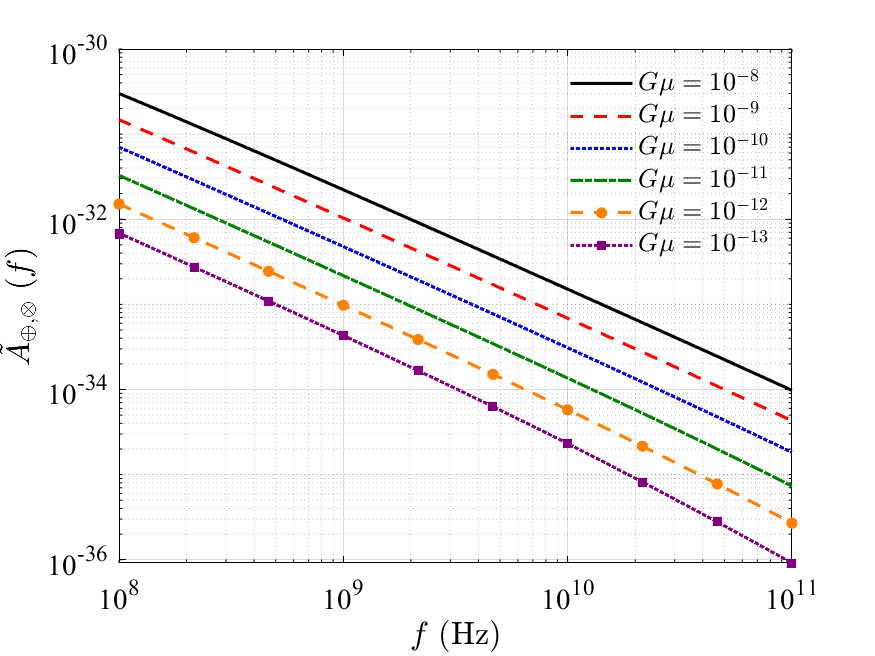}
\caption{Dimensionless amplitude of the SGWB from cosmic strings with different $G\mu$ in the GHz band. The calculation is based on \cite{Aggarwal2021}, using the SGWB spectrum derived in Sect.\ref{sec:2}, with the detailed formulation provided in Sect.\ref{subsec:3.2}.}
\label{fig:2}       
\end{figure}

\begin{figure*}
\centering
  \includegraphics[width=0.4\textwidth]{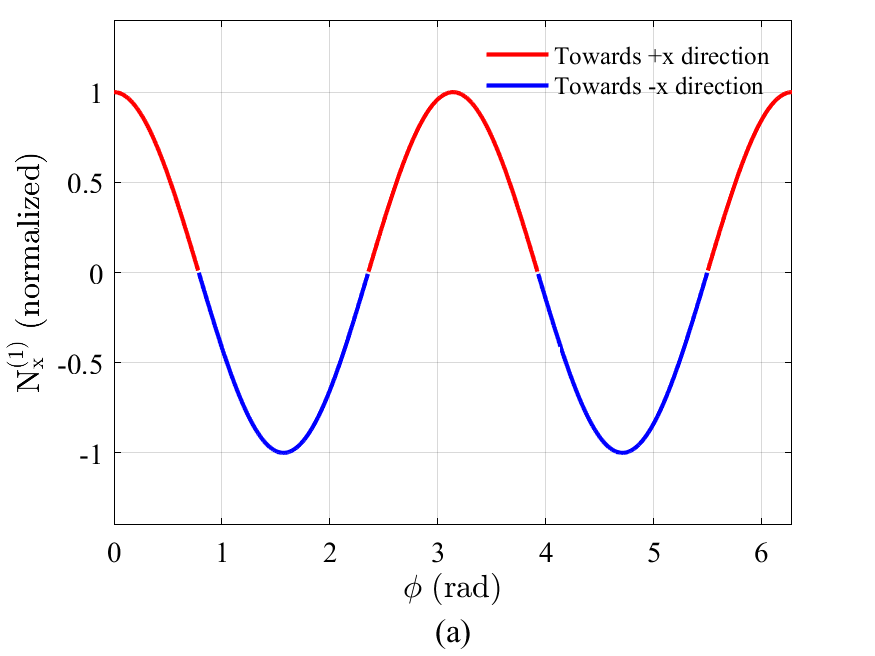}
\includegraphics[width=0.4\textwidth]{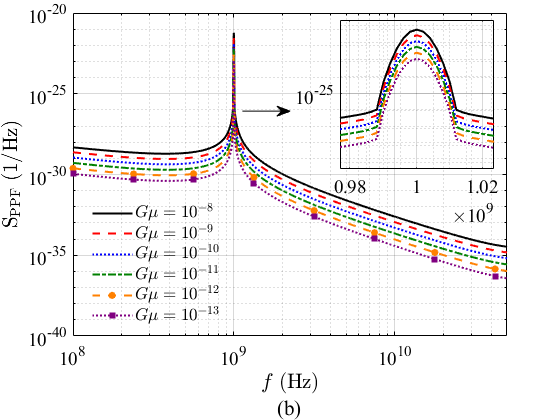}
\caption{(a)The PPFs of the SGWB from cosmic strings vary with $\phi$, here the number of $N^{(1)}_{x}$ is normalized. (b) The energy density spectrum of the PPFs generated by the SGWB from cosmic string in microwave frequency band. Here the related parameter values are consistent with those specified in Table \ref{tab:1}.}
\label{ppfs}       
\end{figure*}

As one kind of SGWB sources, the spectrum of the SGWB from cosmic strings $\Omega_{{\rm{GW}}}(f)$ has the general relationship with the Fourier amplitude $\tilde{h}_{\oplus,\otimes}(f)$ as \cite{Aggarwal2021}
\begin{equation}\label{hf}
<\tilde{h}^{*}_{\oplus,\otimes}(f)\tilde{h}_{\oplus,\otimes}(f')>=\frac{1}{2}\delta(f-f')S_{h}(f),
\end{equation}
\begin{equation}\label{sh}
S_{h}(f)=\frac{3H^{2}_{0}}{4\pi^2}|f|^{-3}\Omega_{{\rm{GW}}}(f),
\end{equation}
where $f=\omega/2\pi$ is the GW frequency, $H_{0}$ describes today's Hubble expansion rate. So the dimensionless amplitude of the SGWB can be derived as $\tilde{A}_{\oplus,\otimes}(f)=\sqrt{fS_{h}(f)}$. On the basis of the spectrum of the SGWB discussed in Sect.\ref{sec:2}, the dimensionless amplitude of the SGWB from cosmic strings with different $G\mu$ in the GHz frequency band is illuminated in Fig.\ref{fig:2}. 
It can be found that at 1GHz the dimensionless amplitude $\tilde{A}_{\oplus,\otimes}<10^{-31}$ and decreases sharply with increasing frequency. This aligns with our theoretical predictions and presents a significant challenge to the detection. So it is necessary to analyze the signal (i.e., PPFs) generated by the SGWB from cosmic strings in the EM resonance system, which can be derived from Eqs.(\ref{h11,h22}),(\ref{h12,h21}), and (\ref{EM11})-(\ref{sh}). Since stochastic gravitational waves arrive from all
directions and Earth-based detectors have fixed orientations while the Earth rotates, it is necessary to consider the averaging effect over spatial positions. Fig. \ref{ppfs}(a) shows how the PPFs of the SGWB from cosmic strings vary with $\phi$. It can be seen that among the full range of $\phi$ (0 to $2\pi$), half of the PPFs values are positive (i.e., the PPFs propagating along the +x direction towards receiver 1), while the other half are negative (i.e., those along the -x direction towards receiver 1' ).  Consequently the spectrum of the PPFs can be further derived as
\begin{eqnarray}\label{FinalNx1}
N^{(1)}_{x}(x,t) = N^{(1)}_{x}|_{\rm{receiver~1}}+N^{(1)}_{x}|_{\rm{receiver~1'}}\nonumber\\
= \frac{1}{\pi}\int_{0}^{\pi}N^{(1)}_{x}(x,t,\phi)d\phi.
\end{eqnarray}

The spectrum of the PPFs after averaging $\phi$ is illuminated as Fig. {\ref{ppfs}}(b), which indicates that the signal (PPFs) exhibits the maximum energy flux near GHz, as the GB's frequency falls within this band, causing an interference-like effect with the first-order perturbation EM field induced by the static magnetic and SGWB interaction in this frequency range. 

\section{The signal processing strategies and sensitivity for the constraints}
\label{sec:4}
In this section, we present a detailed discussion of the detection of the SGWB from the cosmic strings in the EM system. Since the SGWB exhibits isotropic Gaussian random field characteristics, with its energy density spectrum manifesting as random noise in the time domain due to incoherent superposition, it is effective to correlate the outputs of two detectors to detect (or put an upper limit on) a stochastic background signal. Therefore, from the statistical properties of our stochastic gravity-wave background to the signal-to-noise ratio (SNR) calculation will be explained in Sect. \ref{subsec4.1}. Furthermore, the noise in the detection system is an inevitable issue, which will be discussed in Sect. \ref{subsec4.2}. Finally, the sensitivity levels required for the detection are calculated.
\subsection{SNR calculation}\label{subsec4.1}
Supposing that there are two independent such EM resonance systems with outputs in time domain:
\begin{eqnarray}
    \mathcal{D}_{1}(t)=N^{(1)}_{x1}(t)+N_{n1}(t),\nonumber \\
    \mathcal{D}_{2}(t)=N^{(1)}_{x2}(t)+N_{n2}(t),
\end{eqnarray}
where $N^{(1)}_{xi}(t),(i=1,2)$ is the signal (PPFs) in each EM system, $N_{ni}(t),(i=1,2)$ is the number of the noise photons in each EM system. The cross-correlation signal is
\begin{equation}
\mathcal{N}=\int^{T/2}_{-T/2}dt\int^{T/2}_{-T/2}dt'\mathcal{D}_{1}(t)\mathcal{D}_{2}(t')Q(t,t'),
\end{equation}
where $T$ is the observation time, $Q(t,t')$ is a filter function that aims to be dependent only on the time difference $Q(t,t')=Q(t-t')$ rigorously. The optimal choice of filter function $Q(t-t')$ will depend on the locations and orientations of the detectors, as well as on the spectrum of the SGWB and the noise power spectra of the detectors. Since the time-domain waveform of the SGWB can easily resemble and submerged in the combined waveform of various types of noise in detectors, we typically perform the search for the filter function in the frequency domain. Moreover, considering that the noise in different detectors is almost uncorrelated, only the SGWB signal we are interested in is expected to exhibit correlation across data from different detectors. Therefore, we concern the cross-correlation signal in the frequency domain to fix the optimal filter function. Using Fourier transform, $\mathcal{N}$ becomes
\begin{equation}
\tilde{\mathcal{N}}=\int^{\infty}_{-\infty}df\int^{\infty}_{-\infty}df'\delta_{T}(f-f')\tilde{\mathcal{D}}^{*}_{1}(f)\tilde{\mathcal{D}}_{2}(f')\tilde{Q}(f'),
\end{equation}
where 
\begin{equation}
\delta_{T}(f):=\int^{T/2}_{-T/2}dt e^{-{\rm{i}}2\pi ft}=\frac{\sin{(\pi fT)}}{\pi f}
\end{equation}
is a finite time approximation to the Dirac delta function. $\tilde{\mathcal{D}}_{1}(f)$, $\tilde{\mathcal{D}}_{2}(f)$ and $\tilde{Q}(f)$ are the Fourier transforms of $\mathcal{D}_{1}(t)$, $\mathcal{D}_{2}(t)$ and ${Q}(t)$ respectively. In the context of stochastic GW background searches, it is natural to maximize the signal-to-noise ratio, which can be calculated as  
\begin{equation}
{\rm{SNR}}=\frac{<\tilde{\mathcal{N}}>}{\sigma},
\end{equation}
where $<\tilde{\mathcal{N}}>$ represents the mean value of the signal derived as 
\begin{eqnarray}
    <\tilde{\mathcal{N}}>=\int^{\infty}_{-\infty}df\int^{\infty}_{-\infty}<\tilde{N}^{(1)*}_{x1}(f)\tilde{N}^{(1)}_{x2}(f')>\nonumber \\
    \times\delta_{T}(f-f')\tilde{Q}(f')df',
\end{eqnarray}
$\sigma^2=<\tilde{\mathcal{N}}^2>-<\tilde{\mathcal{N}}>^2$ is the variance of the detected data, being related to the noise. According to Eq.(\ref{hij1})-(\ref{Nx1}), the signal $\tilde{N}^{(1)}_{x}(f)$ can be written into a product of the dimensionless amplitude of the SGWB $\tilde{A}_{p}(f),(p=\oplus,\otimes)$ and the EM resonance part $\widetilde{{\rm{EM}}}(f)$ determined by $\hat{B}^{(0)}_{y}$, $\vec{E}^{(0)}$ and $\vec{B}^{(0)}$, yielding
\begin{eqnarray}\label{N}
<\tilde{\mathcal{N}}>=\int^{\infty}_{-\infty}df\int^{\infty}_{-\infty}df'<\tilde{A}^{*}_{p1}(f)\tilde{A}_{p2}(f')>\nonumber \\
  \times<\widetilde{{\rm{EM}}}^{*}_{1}(f)\widetilde{{\rm{EM}}}_{2}(f')> \delta_{T}(f-f')\tilde{Q}(f'),
\end{eqnarray}
setting $\widetilde{{\rm{EM}}}_{1}(f)=\gamma\widetilde{{\rm{EM}}}_{2}(f)$, where $\gamma$ is the overlap reduction function between EM system 1 and 2. Combining with Eq.(\ref{hf}) and (\ref{sh}), Eq.(\ref{N}) can be transformed to be
\begin{equation}
<\tilde{\mathcal{N}}>=\frac{3H^2_{0}}{4\pi^2}T\int^{\infty}_{-\infty}df\frac{\Omega_{{\rm{GW}}}(f)}{f^{2}}\gamma(f)\tilde{Q}(f)|\widetilde{{\rm{EM}}}_{1}(f)|^{2}.
\end{equation}
In order to maximum the SNR, $\tilde{Q}(f)$ should be 
\begin{equation}
\tilde{Q}(f)=\frac{\gamma(f)\Omega_{{\rm{GW}}}(f)\widetilde{{\rm{EM}}}^{2}_{1}(f)}{f^2\tilde{N}^{2}_{n1}(f)\tilde{N}^{2}_{n2}(f)}.
\end{equation}
Thus the maximum filtering SNR of the EM resonance system can be deduced to be 
\begin{eqnarray}
   {\rm{SNR}}^2=(\frac{3H^2_{0}}{4\pi^2})^2T\int^{\infty}_{-\infty}df\frac{\gamma^2(f)\Omega^{2}_{{\rm{GW}}}(f)|\widetilde{EM}_{1}(f)|^{4}}{f^4\tilde{N}^{2}_{n1}(f)\tilde{N}^{2}_{n2}(f)}.
\end{eqnarray}
As a critical point in weak signal detection data processing, a noise characterization is indispensable. Signal detectability can only be achieved when a sufficient SNR threshold is attained. Thus we will discuss the noise spectrum of the EM system $\tilde{N}_{n}(f)$ in the next subsection. 
\subsection{Noise Analysis}\label{subsec4.2}

(1) Background EM noise: As an EM system shown in Fig.\ref{fig:1}, the GB plays an important role for amplifying the energy flux of the perturbed electromagnetic field caused by gravitational waves, but inevitably leads to a background electromagnetic energy flux $\vec{S}^{(0)}$, containing the background transverse perturbative fluxes (BPFs) $N^{(0)}_{x}$ disturbing our signal (PPFs, $N^{(1)}_{x}$). According to Eq.(\ref{Ey0})-(\ref{Bz0}), BPFs can be calculated by
\begin{eqnarray}
    N^{(0)}_{x}(x,t)=\int^{t+\Delta t}_{t}\int\int_{\Delta s}\frac{1}{\mu_{0}\hbar\omega_{e}}<E^{(0)}_{y}B^{(0)}_{z}\nonumber \\
    -E^{(0)}_{z}B^{(0)}_{y}>dydzdt'.
\end{eqnarray}
\begin{figure*}
\centering
\includegraphics[width=0.5\textwidth]{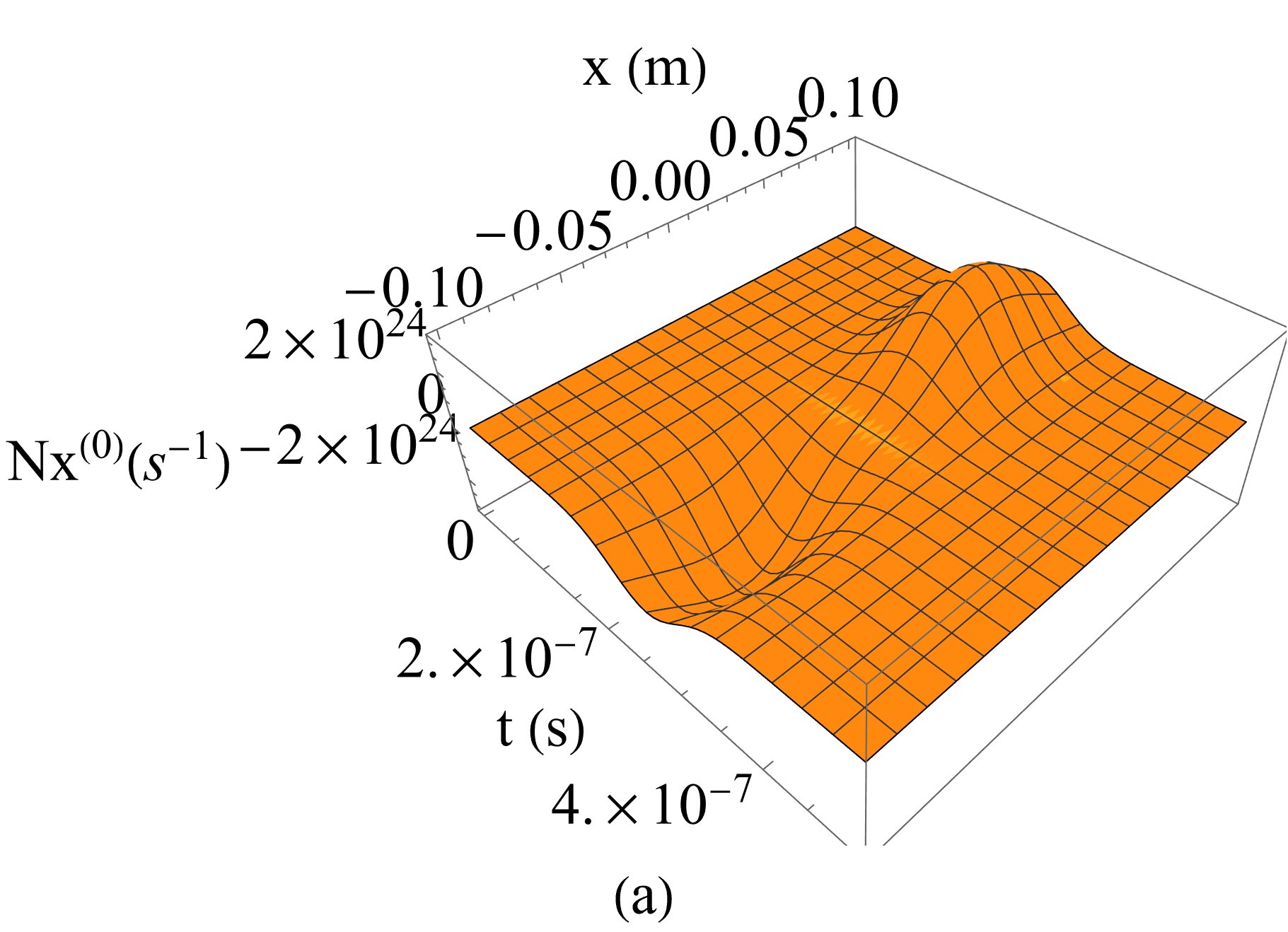}
\includegraphics[width=0.4\textwidth]{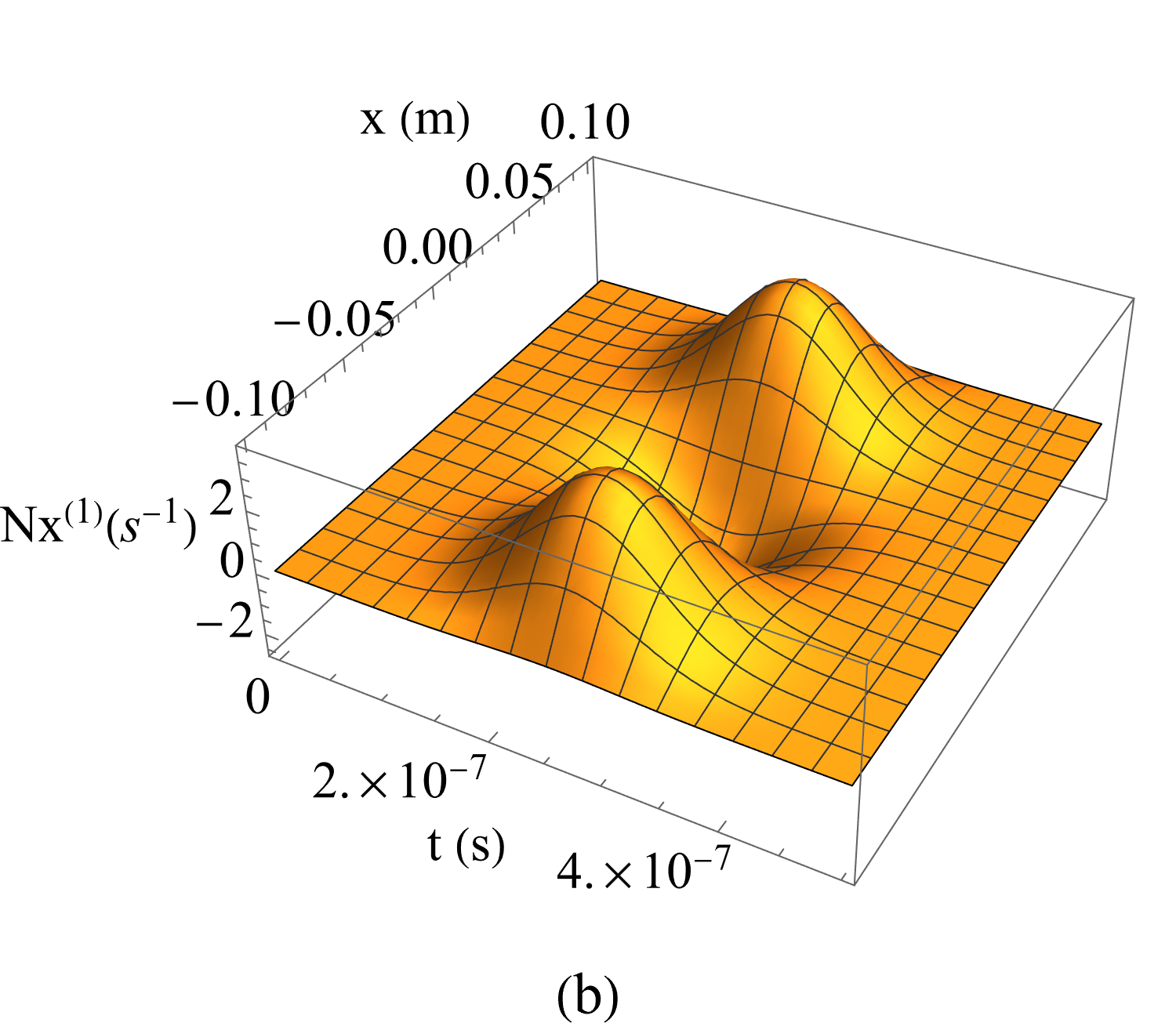}
\includegraphics[width=0.4\textwidth]{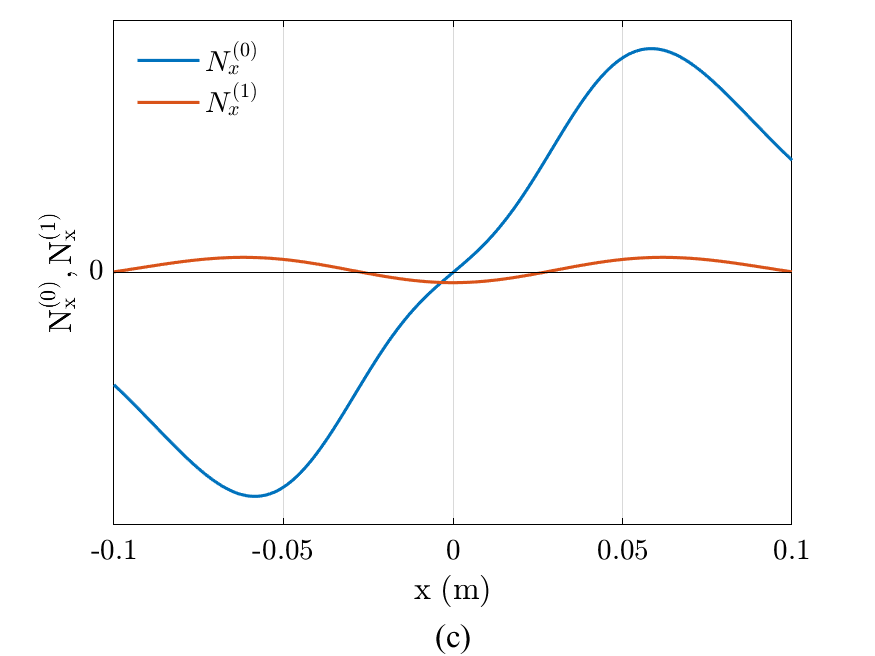}
\includegraphics[width=0.4\textwidth]{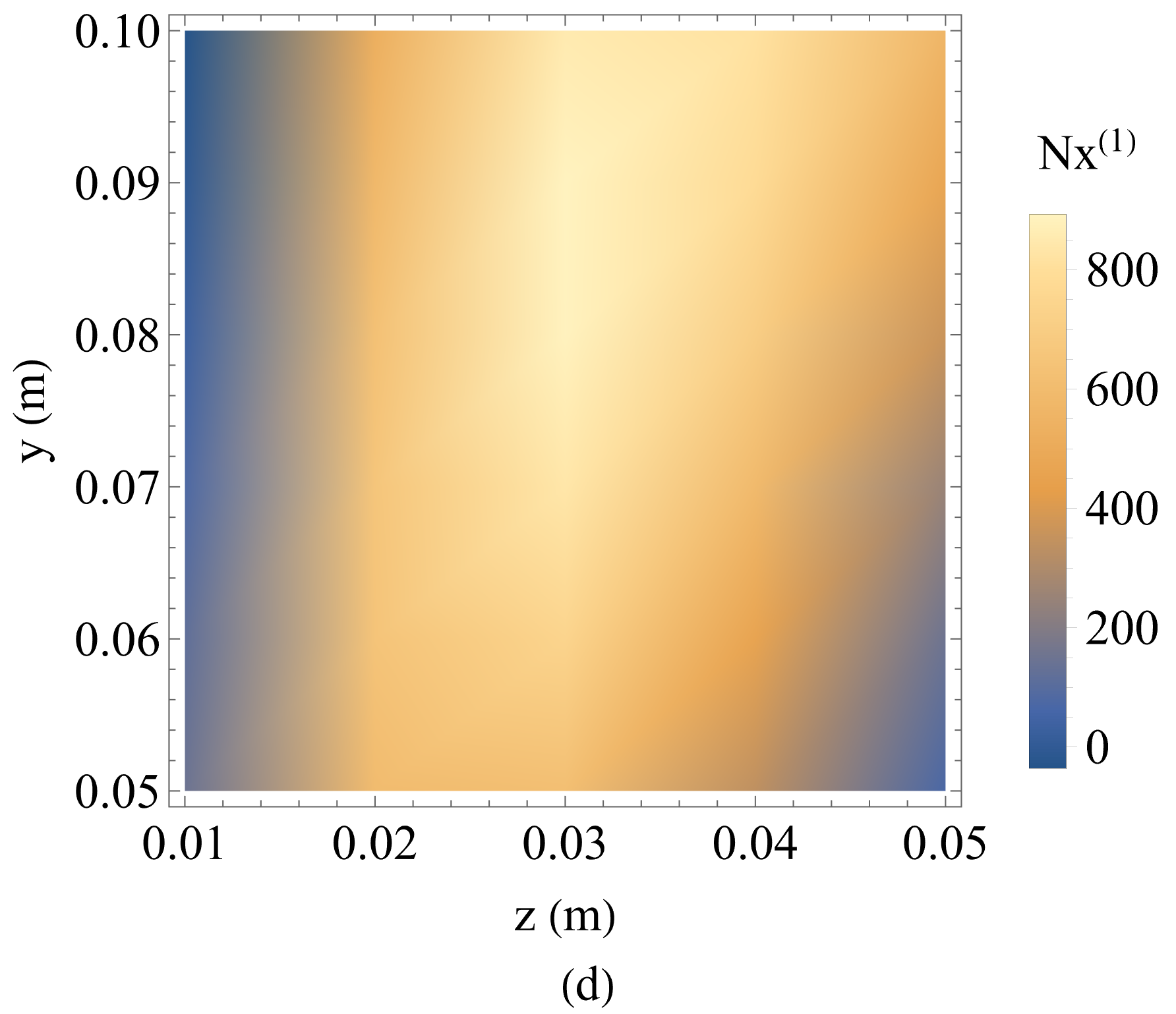}
\caption{(a) Evolution of the background photon flux (BPF, denoted as $N^{(0)}_{x}$) as a function of time and transverse coordinate x. (b) Evolution of the first-order perturbed photon flux (PPF, $N^{(1)}_{x}$) as a function of time and transverse coordinate x. (c) Schematic comparison of background and signal photon fluxes. Due to their significant magnitude difference, both fluxes are rescaled and plotted on the same graph for visual comparison, with modified orders of magnitude on the vertical axis. (d) The PPFs received by receiver 1' at $x=-0.06$m, the detection area is $0.05{\rm{m}}\leq y\leq 0.10{\rm{m}}$ and $0.01{\rm{m}}\leq z\leq 0.05{\rm{m}}$. Here we consider the SGWB with $G\mu=10^{-8}$ and all the parameter values of the EM system are chosen as those in Tab.\ref{tab:1}. The signal accumulation time $\Delta t=1$s.}
\label{BPFvsPPF}       
\end{figure*}
From Fig.\ref{BPFvsPPF} (a) and (b), it can be seen that at the same observation time, BPFs exhibit characteristics of an odd function about the x=0 axis, while PPFs demonstrate features of an even function about the x=0 axis (i.e., in the $x<0$ region, BPFs propagate toward the negative direction of the x-axis, while in the $x>0$ region, BPFs propagate toward the positive direction of the x-axis). That inspires us to detect signals (PPFs) in regions where their propagation directions oppose those of BPFs, thereby enabling effective elimination of the strong background EM noise (BPFs). For optimal selection of the detection regions, Fig.\ref{BPFvsPPF} (c) illustrates the spatial distribution of PPFs and BPFs along the x-axis. As the figure shown, in $-0.10{\rm{m}}<x<-0.03{\rm{m}}$ they propagate in opposite directions, So we choose to put the microwave receiver at $x=-0.06$m where PPFs can be clearly distinguished from the BPFs (cf.Fig.\ref{BPFvsPPF} (d)). Fig.\ref{BPFvsPPF} (d) shows what we can see in the microwave receiver at $x=-0.06$m theoretically, all the received energy fluxes should be from the PPFs (maximum photon number is $\sim$800). Therefore in the total noise, the background EM noise will be excluded. 
\\
(2) Shot noise: The fundamental quantum limit of fluctuations in the detected EM energy fluxes due to the discrete nature of photons. Even in a perfectly EM source, if detecting individual photons (or their statistical effects), there's an inherent randomness in the arrival time of those photons. This randomness causes fluctuations—called shot noise. In this paper, as one kind of systematic noise, the shot noise should contains background shot noise $P_{nb}$ from BPFs and signal shot noise $P_{ns}$ from PPFs. Since the microwave receiver locating at the regions excluded form BPFs (cf. Sect.\ref{subsec4.2}(1)), the corresponding shot noise is also eliminated. Therefore the shot noise here is 
\begin{equation}
P_{\rm{ns}}(f)=hf\sqrt{\tilde{N}^{(1)}_{x}(f)}.
\end{equation}
(3) Thermal noise: Due to the thermal agitation of electrons acting as charge carriers in the microwave receiver, the power of thermal noise $P_{nt}\propto k_{B}T_{t}$ ($k_{B}$ is Boltzmann's constant, $T_{t}$ is the temperature for the system). To reduce the
thermal noise greatly, a cooling system is selected so that the temperature $T_{t}$ satisfies $k_{B}T_{t}\leq\hbar\omega_{e}$ (i.e., $T_{t}\leq48$mK around 1GHz). This condition is satisfied by the temperature for the detector enclosure $T\leq48$mK, which can be conveniently obtained using a common helium-dilution refrigerator so that virtually the number of thermal photons could be significantly suppressed in GHz frequency band. 

At ultra-low temperatures, under the quantum limit ($k_{B}T_{t}\ll\hbar\omega_{e}$), the quantum characteristics exhibited by thermal noise arise from the combined effects of thermal fluctuations and quantum fluctuations, resulting in random signal fluctuations. The power spectral density of the thermal noise is expressed as \cite{RevModPhys.82.1155}: 
\begin{equation}
P_{\rm{nt}}=\frac{hf}{2}\coth(\frac{hf}{2k_{B}T_{t}}).
\end{equation}
(4) Diffraction noise: Diffraction can potentially produce x-axis photons when background EM field is emitted from the GB generator. This is potentially a problem for the system design \cite{Woods2011}. Since the diffraction energy flux density exhibits rapid transverse decay, in our system structure, both the transverse receivers and magnetic aperture positioned outside the background GB field regions, effectively suppressing diffraction noise, the noise from GB diffraction can become negligible. 

Based on the above discussion, the uncorrelated noise components can be root-sum-squared together, so that an expression of the total noise photons is equivalent to
\begin{equation}
\tilde{N}_{n}(f)=\frac{1}{hf}\sqrt{P_{\rm{ns}}^2+P_{\rm{nt}}^2}.
\end{equation}

In addition to the primary noise sources mentioned above, the EM system may occur some other kinds of noise during experimental implementation. These will be incrementally incorporated into our analysis as the hardware system evolves. Nevertheless, the fundamental methodology and framework presented in the paper remain comprehensively applicable.
\subsection{Sensitivity and detectability analysis}\label{subsec4.3}
Based on the comprehensive analysis presented above, we numerically characterize the system's fundamental sensitivity and estimate detectable signal strength. 
\begin{figure*}
\centering
\includegraphics[width=0.4\textwidth]{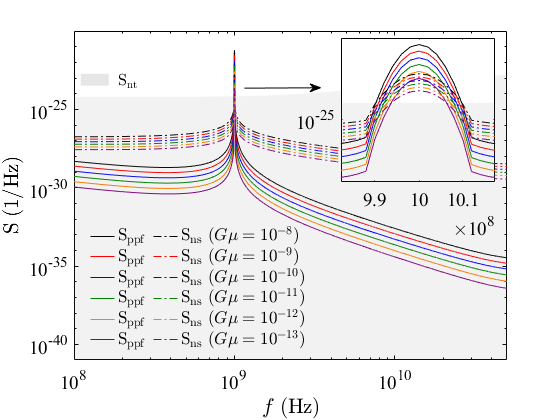}
\includegraphics[width=0.4\textwidth]{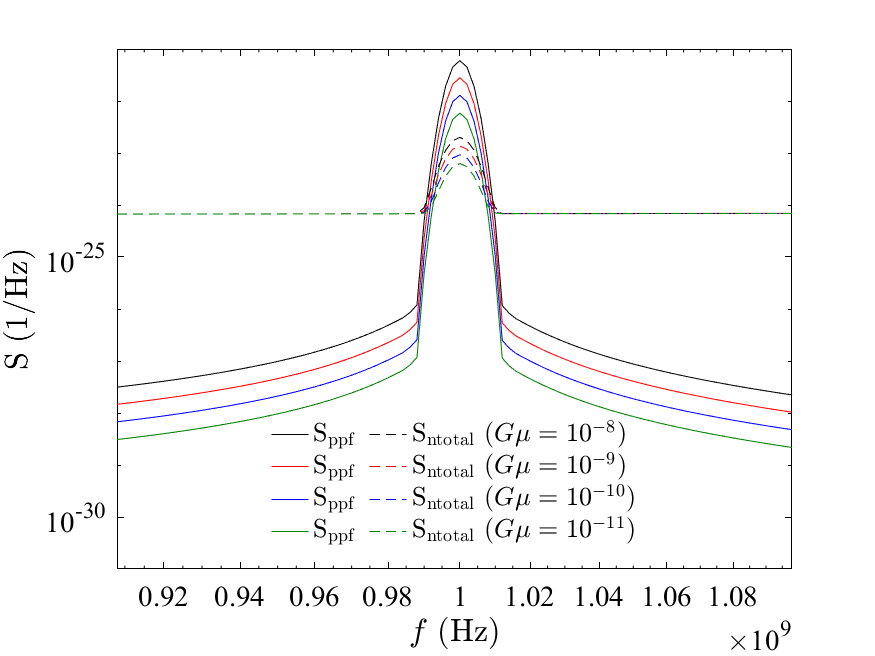}
\caption{\textbf{Left-panel:}the spectrum of energy density for signal (S$_{{\rm{ppf}}}$), shot noise (S$_{{\rm{ns}}}$) and thermal noise (S$_{{\rm{nt}}}$). Different colors indicate the energy density with different $G\mu$. All solid lines represent the energy density spectra of the signal PPF, whose intensity varies due to the different values of $G\mu$, while their shapes remain similar. All dashed lines represent the energy density spectra of the shot noise, which originates from the scattering of the signal (PPF); therefore, their intensities also vary with $G\mu$, but their shapes are similar. The thermal noise caused by the ambient temperature is shown as a shaded region so as not to interfere with the shot noise curves. \textbf{Right-panel:} the sensitivity of the EM system (S$_{{\rm{ntotal}}}$) and the detectable signal energy density spectrum. Here $T_{t}=45$mk}
\label{signalvsnoise}       
\end{figure*}
Then from the spectrum of signal and noise energy densities in Fig.\ref{signalvsnoise}, it is evident that the
SGWB signal exhibits a pronounced peak energy density around the central frequency of the GB, marking the primary detection region. As the dimensionless string tension $G\mu$ increases, the peak energy density grows accordingly. For the maximum value of string tension $G\mu=10^{-8}$, the SNR reaches $270$ (cf. Tab. \ref{tab:2}). The threshold of SNR chosen to be 20 with false alarm rate $\mathcal{P}_{{\rm{FA}}}=0.001$, the EM resonance system is able to achieve a SNR$\geq$20 for the SGWB with tension $G\mu=10^{-11}, 10^{-10}, 10^{-9}, 10^{-8}$ around 1 GHz.
\begin{table}
\centering
\caption{Signal-to-Noise Ratio (SNR) and the corresponding ${\rm{PPFs}}$ (i.e.,$N^{(1)}_{x}$) for different values of the dimensionless string tension $G\mu$. Here the observation time $T=1$s, $\gamma=1/2$. }
\label{tab:2}
\begin{tabular}{lll}
\hline\noalign{\smallskip}
$G\mu$ & SNR & PPFs (s$^{-1}$)  \\
\noalign{\smallskip}\hline\noalign{\smallskip}
$10^{-8}$  & 273.99 & 893.00 \\
$10^{-9}$  & 120.50 & 416.44 \\
$10^{-10}$ & 51.04 & 191.81 \\
$10^{-11}$ & 20.64 & 87.39 \\
$10^{-12}$ & 7.86 & 39.27 \\
$10^{-13}$ & 2.75 & 17.29 \\
\noalign{\smallskip}\hline
\end{tabular}
\end{table}
That means through the EM resonance system it is possible to detect the SGWB with cosmic string tension $G\mu\geq 10^{-11}$, or constrain the cosmic string tension $G\mu<10^{-11}$ in GHz frequency band. 

\section{Summary and prospects}\label{sec:5}

The fundamental concept of our HFGW EM resonance system involves utilizing the fluctuating electromagnetic components along the longitudinal direction of a GB to amplify weak second order ($\sim h^{2}$) electromagnetic fields generated through GW and static magnetic field coupling (derived from solving electrodynamic equations in curved spacetime Eqs.(\ref{curveeq1}), (\ref{curveeq2})). This amplification creates detectable first-order energy fluxes.

In the microwave band, the dimensionless amplitude of gravitational waves is extremely small. Even after Gaussian beam amplification, the resulting photon flux remains susceptible to being overwhelmed by the GB's intrinsic background photons. Through analyzing their distinct transverse decay rates and propagation directions (cf. Sect.\ref{subsec4.2}), we identified an optimal detection region. Without the GB's perturbing intrinsic background photons (i.e., where PPFs and BPFs propagating in the opposite direction), we successfully isolated first-order transverse perturbed electromagnetic energy flux (i.e., PPF) outside the beam waist in $0.997$–$1.003$ GHz. The PPF imprints of gravitational wave effects, enabling us to establish constraints on the cosmic string tension parameter $G\mu\simeq 10^{-11}$ with a SNR threshold of 20.

While space-based gravitational wave detectors may probe cosmic string SGWB with $G\mu\simeq 10^{-17}$–$10^{-18}$ \cite{wang2023probing,PhysRevD.109.063520}, our microwave band approach exhibits weaker constraints on $G\mu$. Nevertheless, it provides complementary observational capabilities. Certainly , there are several aspects require refinement. In terms of theoretical aspects, we can incorporate other gravitational wave sources existing in the microwave band as detection targets, such as first-order phase transitions and mergers of primordial black holes \cite{li2017test}. In experimental aspects, there are shielding against environmental microwave electromagnetic background, spatial homogeneity optimization of static magnetic fields, a more detailed overlap reduction function $\gamma(f)$ and so on. These represent critical directions for follow-up studies building upon our current framework. This work establishes an experimental foundation for future GW observations in microwave band.

\paragraph{{\rm{\textbf{ACKNOWLEDGMENTS}}}}

This work was supported by the National Key Research and Development Program of China (Grant No. 2021YFC2203004), the Fundamental Research Funds for the Central Universities Project (Grant No. 2024IAIS-ZD009), the National Natural Science Foundation of China (Grants No. 12575072, No. 12347101, No. 12505057, and No. 12505058), the Natural Science Foundation of Chongqing (Grant No. CSTB2023NSCQ-MSX0103), and the Brazilian agencies FAPESQ-PB.

%




\end{document}